\documentclass[11pt]{article}

\usepackage[a4paper, total={6.5in, 8.5in}]{geometry}

\usepackage[utf8]{inputenc}
\usepackage[english]{babel}
\usepackage[T1]{fontenc}

\usepackage{xcolor}
\usepackage{graphicx}
\usepackage{amsfonts, amsmath}
\usepackage{bbold}
\usepackage{framed}
\usepackage{hyperref}
\usepackage{listings}

\definecolor{shadecolor}{rgb}{0.94, 0.97, 1.0}

\newcommand{\lil}{\lstinline[basicstyle=\ttfamily]}

\newcommand{\be}[1]{\begin{equation}\label{#1}}
\newcommand{\ee}{\end{equation}}

\newcommand{\ket}[1]{| #1 \rangle}

\newcommand{\Hint}{$\mathcal{H}_{\textrm{int}}$}

\begin{document}

\title{Qadence: a differentiable interface for digital-analog programs}

\author{
  Dominik Seitz,
  Niklas Heim,
  João P. Moutinho,
  Roland Guichard,\\
  Vytautas Abramavicius,
  Aleksander Wennersteen,
  Gert-Jan Both,
  Anton Quelle,\\
  Caroline de Groot,
  Gergana V. Velikova,
  Vincent E. Elfving,
  Mario Dagrada\\
  \\
  PASQAL, 7 rue L\'eonard de Vinci, 93100 Massy, France
}
\date{}















\maketitle

\begin{abstract}
  Digital-analog quantum computing (DAQC) is an alternative paradigm for universal quantum computation combining digital single-qubit gates
  with global analog operations acting on a register of interacting qubits. 
  Currently, no available open-source software is tailored to express, differentiate, and execute programs within the DAQC paradigm. 
  In this work, we address this shortfall by presenting Qadence, a high-level programming interface for building complex digital-analog quantum programs developed at Pasqal. Thanks to its flexible interface, native differentiability, and focus on real-device execution, Qadence aims at advancing research on variational quantum 
  algorithms built for native DAQC platforms such as Rydberg atom arrays.
\end{abstract}

\section{Introduction}\label{sec:intro}

Most of the progress towards scalable quantum computation has been focused on the digital quantum computing paradigm for NISQ\footnote{Noisy Intermediate Scale Quantum.} devices. Naturally, the development of high-level quantum programming languages has been geared towards writing programs within this paradigm. Digital quantum programs are typically written within the so-called \emph{quantum circuit} model as a composition of operations, or \textit{gates}, acting on a limited subset of non-interacting qubits. Several open-source quantum software development kits are now available for this purpose such as 
Qiskit \cite{qiskit}, Cirq \cite{cirq}, and Pennylane \cite{pennylane}.

Nevertheless, few quantum computational paradigms constitute today valid alternatives to the digital one for programming near-term devices. Among them, one of the most promising is Digital-Analog Quantum Computing (DAQC). DAQC was introduced in Parra-Rodriguez \textit{et al.} \cite{solano2020} based on the seminal work by Dodd \textit{et al.} \cite{dodd2002} where it has been shown to be universal. Several common quantum algorithms, such as the Quantum Fourier Transform (QFT) \cite{daqft2020} or
the quantum approximate optimization algorithm (QAOA) \cite{qaoa2014}, have been successfully cast into their digital-analog counterparts. The most salient differences 
between DAQC programs and digital ones are the following:
\begin{itemize}
    \item A global interaction Hamiltonian, \Hint, used as a primary 
    computational resource 
    for a given register of qubits, forming complex \emph{analog operations}. This 
    avoids isolating interactions for specific multi-qubit digital gates.
    \item The completion of analog operations by parameterized local single-qubit rotations.
\end{itemize}
As such, the concrete form of \Hint\;is strongly dependent on the physical realization of the digital-analog device. Furthermore, while particular systems
allow for \Hint\;to be easily switched on or off, others operate more naturally with an \textit{always-on} interaction. For certain algorithms, the customizable \Hint\; can also be directly exploited as a computational resource, for instance, as a native entangling layer in variational circuits \cite{hea}.

Recently, Rydberg atom arrays have surged as a promising platform for the development of scalable quantum computation, with devices expected to reach thousands of physical qubits in the near term \cite{loic2020}. Besides fully digital operations, these devices also excel at operating in analog mode with the ability to perform global pulses, acting on customizable qubit register topologies \cite{scholl2021quantum}, making them perfectly suited candidates for the physical realization of DAQC programs. In the past few years, interfaces for programming Rydberg atom arrays have focused on pulse-level programming \cite{pulser, bloqade2023quera}. At this level of abstraction, the user is provided with the necessary functionalities 
to fine-tune and control all technical details required for a neutral atom experiment. Higher-level programming interfaces therefore need to stretch between low-level hardware-specific requirements for near-term quantum algorithms and high-level abstractions. In the purely analog mode, a recent attempt aimed at developing a \textit{Hamiltonian modeling language} \cite{simuq}, purposefully abstracting away device programming details to the Hamiltonian itself. However, no programming interface exists facilitating an easy user uptake coming from the more widespread digital paradigm into the DAQC world.

To bridge this gap, we present Qadence \cite{qadence}, a high-level programming interface for efficiently writing, differentiating, and executing DAQC programs on compatible devices such as Rydberg atom arrays developed at Infleqtion, Planqc, Atom Computing, QuEra and Pasqal. In Section \ref{sec:api} we present the main building blocks of the Qadence API. Qadence includes a flexible block system approach to composing quantum operations (Section \ref{subsec:blocks}) as well as the definition and arbitrary composition of variational and feature parameters (Section \ref{subsec:parameters}). It allows the creation of qubit registers with arbitrary topology for digital-analog computations, and respective unification with the block system into a quantum circuit (Section \ref{subsec:register_circuit}). Furthermore, Qadence supports the usage of differentiable backends with various differentiation modes to support the execution of variational programs (Section \ref{subsec:exec}). All of these building blocks are unified in the Quantum Model interface (Section \ref{section:quantum_model}). In Section \ref{sec:dap} we show how Qadence simplifies the creation of digital-analog quantum programs, both in compilation to pre-defined Hamiltonians and in the creation and manipulation of arbitrary Hamiltonian operations. In Section \ref{sec:apps}, we show some simple applications in quantum machine learning (QML) and digital-analog programming, showcasing the full breadth of Qadence capabilities. Finally, conclusions and perspectives are drawn in Sections \ref{sec:concl} and \ref{sec:future}.

\section{Qadence overview}\label{sec:api}

\begin{figure}[t]
    \centering
    \includegraphics[width=0.7\linewidth]{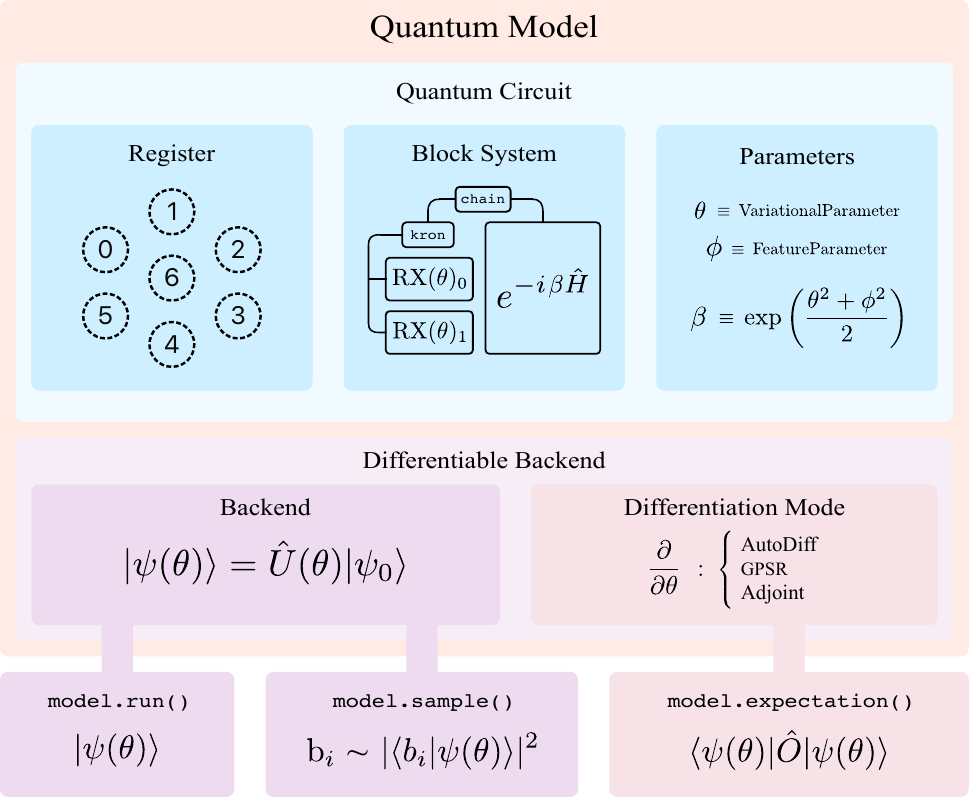}
    \caption{Qadence high-level architecture diagram together with main components and their relations.
    Programs are primary focused on using the \lil{QuantumModel} object, which ties a composed 
    \lil{QuantumCircuit} with a \lil{DifferentiableBackend}, useful for machine learning (ML) workloads execution with fully integrated differentiation engines. Here, Qadence builds on 
    top of PyTorch \cite{pytorch} for automatic differentiation of statevector simulators, while also
    providing a native integration of generalized parameter-shift rules \cite{gpsr2021} for general differentiability of quantum variational programs.
    }
    \label{fig:arch}
\end{figure}

In this section, we present Qadence's general architecture and user API. The circuit construction features of Qadence are designed to work for arbitrary digital platforms, with a strong focus on digital-analog devices. As a basic design tenet, Qadence adopts a fully functional approach to the low-level components while offering an object-oriented and flexible user API. The overall architecture of the library is shown in Fig.~\ref{fig:arch}.

\subsection{A flexible block system}\label{subsec:blocks}

\begin{figure}[t]
    \centering
    \includegraphics[width=\linewidth]{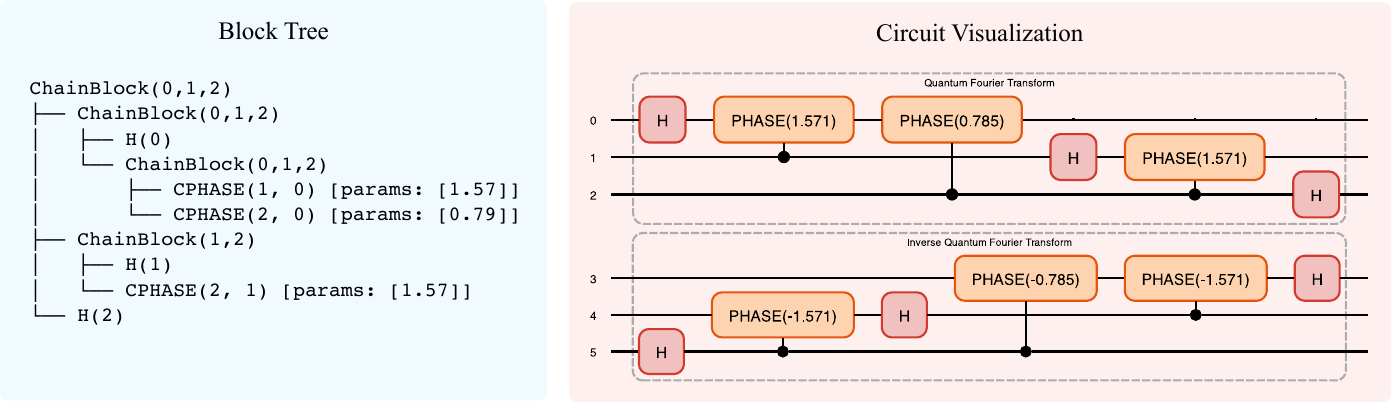}
    \caption{Qadence uses a flexible block system focused on operations modularity, heavily inspired by the Yao package and Quipper \cite{yao, quipper}. By default, \lil{print(block)} shows the
    respective block tree in the console output, as exemplified on the left-hand side for a single QFT block. Blocks can also be tagged with \lil{tag(block, "tag name")} and the circuit visualized with the \lil{display} function available in the \lil{qadence.draw} module.
    }
    \label{fig:blocks}
\end{figure}

Essential to the extensibility and flexibility of Qadence is its core
block-based programming interface, heavily inspired by the Yao package \cite{yao} and Quipper \cite{quipper} (see Fig.~\ref{fig:blocks}). We consider a block to be a composable representation of a quantum operation, such as a single or multi-qubit digital gate, a Hamiltonian or the respective evolution operation, or any arbitrary composition of operations. Composing blocks can be done with:
\begin{itemize}
    \item \lil{chain(A, B)} or \lil{A * B}, composing sequentially through normal operator multiplication;
    \item \lil{kron(A, B)} or \lil{A @ B}, composing in parallel through the Kronecker product;
    \item \lil{add(A, B)} or \lil{A + B}, directly adding operators together.
\end{itemize}
These composition functions automatically build the qubit support of the composed block from each individual block's one. As an example, consider the simple code\footnote{As exemplified in this code snippet, user-facing functions in Qadence are typically accessible from the main namespace, and an alternative suggested use is to do [\lil{import qadence as qd}]. Nevertheless, in the Qadence documentation, imports are explicitly defined, but omitted in the remainder of the examples for the sake of readability.} below defining a digital QFT on an arbitrary qubit support.

\begin{shaded*}
\begin{lstlisting}[caption=Digital Quantum Fourier Transform, label={lst:qft}, language=python]
from qadence import *

def cphases(qs: tuple, l: int):
    return chain(CPHASE(qs[j], qs[l], PI/2**(j-l)) for j in range(l+1, len(qs)))

def QFT(qs: tuple):
    return chain(H(qs[l]) * cphases(qs, l) for l in range(len(qs)))

# Easily compose a QFT and its inverse
qft_block = kron(QFT((0, 1, 2)), QFT((3, 4, 5)).dagger())
\end{lstlisting}
\end{shaded*}

This block system flexibility greatly facilitates syntactic compositions. As quantum computing advances and the number of qubits scales, low-level blocks requiring a manual specification of individual operations and their targets can be abstracted away and users will interact only with higher-level block constructors for more efficient writing of quantum programs.

\subsection{Symbolic expressions as parameters}\label{subsec:parameters}
Qadence relies heavily on the symbolic algebra package Sympy \cite{sympy} for defining parameters and their arbitrary transformations or compositions. It discriminates between three \lil{Parameter} types:
\begin{itemize}
    \item \emph{Fixed Parameter}: constant, with a fixed non-trainable value (e.g. $\pi/2$).
    \item \emph{Variational Parameter}: trainable, to be further optimized.
    \item \emph{Feature Parameter}: non-trainable, requiring an input value, usually used for encoding classical data into qubit rotations.
\end{itemize}

This offers a convenient interface to build complex symbolic expressions to, for instance, parametrize feature maps, define variational ans{\"a}tze circuits, or fully parameterize Hermitian operators compatible with Hamiltonian evolution and observable measurements. A defined parameter can also be re-used in several distinct blocks throughout the program (see Code Sample \ref{lst:expressions}).

\begin{shaded*}
\begin{lstlisting}[caption=Arbitrary expressions as parameters, label={lst:expressions}, language=python]
# Defining a Feature and Variational parameter
theta, phi = VariationalParameter("theta"), FeatureParameter("phi")

# Alternative syntax
theta, phi = Parameter("theta", trainable = True), Parameter("phi", trainable = False)

# Arbitrarily compose parameters with sympy
expr = sympy.acos((theta + phi)) * PI

gate = RX(0, expr) # Use as unitary gate arguments
h_op = expr * (X(0) + Y(0)) # Or as scaling coefficients for Hermitian operators
\end{lstlisting}
\end{shaded*}

\subsection{Qubit register and circuit composition}\label{subsec:register_circuit}

\begin{figure}[t]
    \centering
    \includegraphics[width=0.9\linewidth]{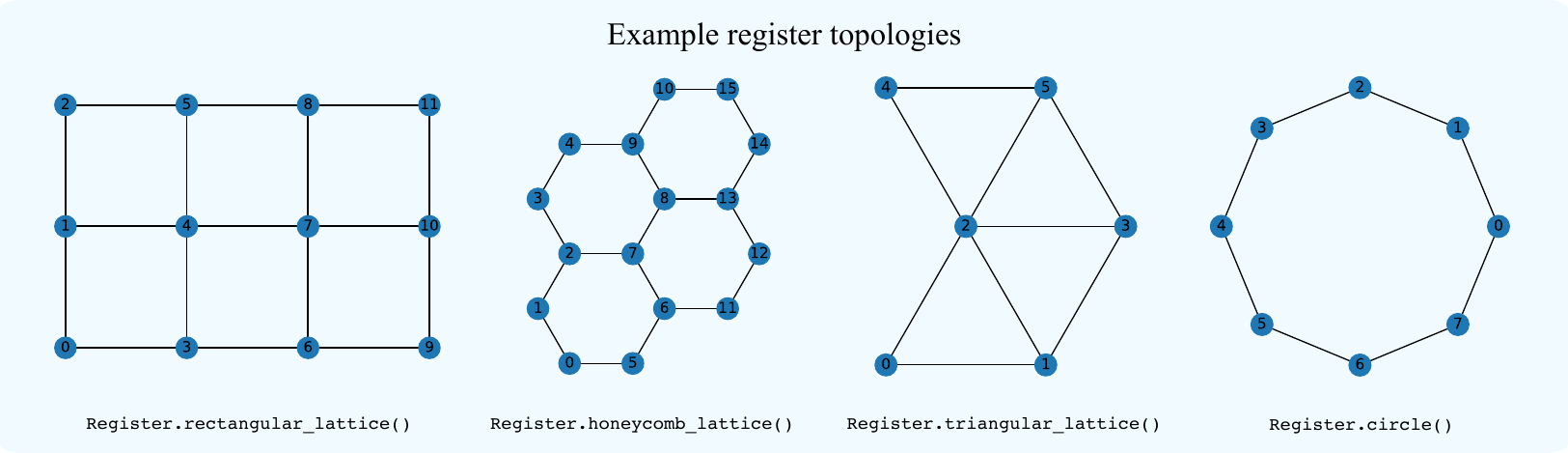}
    \caption{Registers in Qadence encode a NetworkX graph to represent the topology, where each qubit coordinates are node properties. Typically, register constructors will define an edge for each pair of neighboring qubits accessible with the \lil{register.edges} method. While these edges can be used to represent interactions in abstract Hamiltonians, they are not necessarily representative of the interaction topology in real qubit systems. Instead, all node pairs and respective distances can be accessed with the \lil{register.all_node_pairs} and \lil{register.distances} methods. These are convenient tools for the creation of arbitrary \Hint, which sum over the complete graph pairs with an interaction strength that decays with the distance (exemplified in Sample \ref{lst:arb-ham}).}
    \label{fig:register}
\end{figure}

Besides block compositions as operations, a quantum program also requires a resource register of qubits to act on. In Qadence, the \lil{Register} is a NetworkX \cite{networkx} graph object containing the total number of qubits and their spatial coordinates. It is necessary to determine interaction strengths in the creation of \Hint\;for particular DAQC platforms (e.g. Rydberg atom arrays).

\begin{shaded*}
\begin{lstlisting}[caption=Register of qubits., label={lst:register}, language=python]
reg = Register.line(3) # A simple line register of 3 qubits

# Use other topologies, and set the spacing between qubits
reg = Register.from_coordinates([(0.0, 0.0), (0.0, 1.0), (0.0, 2.0)])
reg = Register.circle(n_qubits = 10, spacing = 10)
reg = Register.triangular_lattice(n_cells_row = 2, n_cells_col = 2)
\end{lstlisting} 
\end{shaded*}

\noindent Finally, the \lil{QuantumCircuit} brings together both a resource register and a composite block of operations for the program to be executed. For digital programs assuming all-to-all qubit connectivity, where only the number of qubits is needed, an integer can be passed directly to automatically create a \lil{Register} instance.

\begin{shaded*}
\begin{lstlisting}[caption=Quantum circuit definition, label={lst:qcirc}, language=python]
circuit = QuantumCircuit(reg, blocks) # Circuit using a pre-defined register
circuit = QuantumCircuit(n_qubits, blocks) # Circuit using the number of qubits directly
\end{lstlisting} 
\end{shaded*}

\subsection{Backend execution and differentiability}
\label{subsec:exec}

A \lil{QuantumCircuit} instance as defined in the previous section still remains an abstract object. Concretization and execution requires a \emph{backend} (see Fig.~\ref{fig:arch}), which can be either an actual physical quantum device such as a Rydberg atomic processor, or a numerical simulator. In the following, we focus on numerical simulators.

\subsubsection{Backend simulators}\label{subsubsec:backends}

Qadence is compatible with multiple numerical simulation backends, including some custom-built ones for its use:

\begin{itemize}
    \item \textbf{PyQTorch} \cite{pyqtorch}: a differentiable and exact statevector simulator written in PyTorch for gate-based quantum computing operations.
    \item \textbf{Pulser} \cite{pulser}: a pulse-level programming interface for neutral atom devices supporting time-dependent pulse simulation and real-device execution.
    \item \textbf{Emu-C} \cite{anton_2024_emuc}: a tensor network simulator using PyTorch with built-in differentiation. It supports several contraction strategies and singular-value decomposition (SVD) based truncation. 
    \item \textbf{Horqrux} \cite{horqrux}: also a differentiable and 
    exact statevector simulator. Horqrux is, however, written in 
    JAX \cite{jax} and is currently experimental.    
\end{itemize}

In the following, we provide some details on two backends developed specifically for Qadence: PyQTorch and Emu-C. \footnote{Note that, unlike PyQTorch, Emu-C is currently closed-source, with a planned open-sourcing during Q2 of 2024.}
Qadence is also executable through Pasqal's cloud platform, where Emu-C, PyQTorch, and Pulser are selectable backends along the pulse-level tensor network backend developed in \cite{emutn_on_cloud}.

\paragraph{PyQTorch}
\label{subsubsec:pyqtorch}

A lean (less than 1000 lines of code) numerical statevector 
simulator written on top of the widespread PyTorch  
deep learning framework. 
PyQTorch provides all the common digital quantum operations, including
parametrized ones, and allows time-independent Hamiltonian exponentiation for analog operations. In PyQTorch, each quantum operation is implemented as a separate PyTorch \lil{nn.Module} instance, applying the corresponding matrix to the quantum state, defined as a dense PyTorch tensor. This allows to seamlessly combine quantum operations into circuits, and ensures native support for automatic differentiation (AD) with the PyTorch \lil{autograd} engine. AD support extends to most quantum routines except bitstring sampling.

\paragraph{Emu-C}
\label{subsubsec:emu_c}

A numerical simulator based on tensor networks (TN),
more specifically Matrix Product State (MPS) \cite{orus_2014_tns},
that supports AD using PyTorch.
In the MPS formalism we represent the large quantum state tensor $\ket{\psi}$
as a 1-dimensional chain of single-site tensors.

Emu-C executes the abstract circuit representation by translating Qadence blocks into TN operators.
Factorized operators and MPS are generated via sequential Schmidt decomposition 
$\ket{\psi} = \sum_i^D \lambda_i \ket{\text{Left}_i} \otimes \ket{\text{Right}_i}$ \cite{hand_wavy, tensorly_quantum}, such that the resulting tensors are single-sited.
The dimension $D$ dictates the efficiency of the TN representation, and also encodes the maximum entanglement between left and right parts.
MPS are further truncated through singular-value decomposition (SVD), reducing $D$, thus allowing control over memory requirements. Finally, we perform the desired calculation through network contraction over joined tensor indices. Emu-C allows both approximate and exact calculation of the circuit via several contraction strategies. Exact calculations can be performed in a ``naive'', or layer-based, manner or using state of the art hyper-graph partitioning \cite{cotengra}.

\subsubsection{Differentiable Backend}

Qadence supports a \lil{DifferentiableBackend} which, in conjunction with a quantum \lil{Backend} instance, allows users to create variational programs with built-in differentiation working on both emulated and real quantum devices. Derivatives of quantum circuits outputs with respect to feature and variational parameters can be computed via three differentiation modes: automatic differentiation (AD) from seamless integration with PyTorch and JAX engines, adjoint differentiation \cite{adjoint} and generalized parameter-shift rule (GPSR) \cite{gpsr2021}. Furthermore, gradient-based optimizers can also be readily used for faster convergence of the variational procedure. Details on the \lil{DifferentiableBackend} low-level API and the available differentiation modes invocation in Qadence can be found Appendix \ref{appendix:diff}.


\subsubsection{Differentiation Modes}

\paragraph{Automatic differentiation (AD)}
Automatic differentiation comprises methods for evaluating the exact derivatives of numerical functions that allows for the training of large and complex machine learning models via backpropagation \cite{baydin2017}. The reverse-mode approach of calculating gradients is the dominating paradigm in ML learning frameworks such as PyTorch and JAX, with the evaluation of higher-order derivatives relying on the sequential application of the AD operator \cite{Griewank2008}. AD is supported by the pyqtorch, emu-c, and horqrux backends.




\paragraph{Adjoint Differentiation (ADJOINT)}
The pyqtorch backend also provides an implementation of \emph{adjoint differentiation} \cite{adjoint}. Similarly to AD, adjoint differentiation is not device-compatible. However, by exploiting the intrinsic reversibility of quantum operations, it provides an efficient alternative to standard reverse-mode AD, allowing the memory consumption to scale only with the number of qubits, and not with the number of parameters or the depth of the circuit. The algorithm described in Tyson et al. \cite{adjoint} and therefore also the implementation of adjoint differentiation in Qadence, currently supports first-order derivatives for digital gates.

\paragraph{Generalized parameter-shift rule (GPSR)}

Qadence includes an implementation of the generalized parameter-shift rule from Kyriienko \textit{et al.} \cite{gpsr2021}, which enables high-order differentiation for any numerical backend that
provides an \lil{expectation} method, and is also realizable on physical devices. 
It allows indirect differentiation of arbitrary generators of quantum operations with respect to a parameter $x$, by defining the expectation value of some cost operator $\hat{C}$ :
\be{eq:gpsr_f}
f(x) = \left\langle 0\right|\hat{U}^{\dagger}(x)\hat{C}\hat{U}(x)\left|0\right\rangle.
\ee
$\hat{U}(x)=\exp(-i\frac{x}{2}\hat{G})$ is the quantum evolution operator generated by $\hat{G}$, which holds the structure of the underlying quantum circuit. From the eigenvalue spectrum of $\hat{G}$, it is possible to calculate the full set of corresponding unique non-zero spectral gaps $\left\{ \Delta_s\right\}$. The final expression for the derivative of $f(x)$ is then given by the following expression:
\begin{equation}
\frac{df\left(x\right)}{dx}=\overset{S}{\underset{s=1}{\sum}}\Delta_{s}R_{s},
\end{equation}
where $S=|\left\{ \Delta_s\right\}|$ and $R_s$ are real solutions of a system of linear equations for shifted parameters\footnote{\href{https://pasqal-io.github.io/qadence/latest/advanced_tutorials/differentiability/}{\lil{https://pasqal-io.github.io/qadence/latest/advanced_tutorials/differentiability/}}}. GPSR is backend-agnostic.

\subsection{The quantum model interface}
\label{section:quantum_model}

The \lil{QuantumModel} class integrates all the aforementioned components allowing a user to build their quantum programs with a straightforward and user-friendly interface. It inherits from the \lil{torch.nn.Module} class, for seamless composition with conventional neural network models, on top of supporting all the standard built-in and external PyTorch-based libraries for the programmer's convenience. The \lil{QuantumModel} exposes the following methods for program execution:
\begin{itemize}
    \item \lil{run}: Sequentially apply a series of quantum operations defined in a circuit to an initial state. Returns an array of shape $(batchsize, 2^{N})$ ($N$ represents the number of qubits in the full system) when using a state vector simulator as a backend.
    \item \lil{sample}: Measure the circuit in the computational basis. Returns a dictionary mapping bitstrings to integers representing the number of counts.
    \item \lil{expectation}: Compute the expectation value of a given observable following a circuit. Returns an array of shape $(batchsize, N_\text{obs})$, where $N_\text{obs}$ represents the number of observables measured.
\end{itemize}

\begin{shaded*}
\begin{lstlisting}[caption=Quantum model interface and differentiation., label={lst:qm}, language=python]
# We start with a simple block composition using some convenience constructors
n_qubits = 4
fm = feature_map(n_qubits, param="phi") # Different feature maps are directly available
ansatz = hea(n_qubits, depth = 1) # Standard digital Hardware-Efficient Ansatz
circ = QuantumCircuit(n_qubits, fm * ansatz)

# Observable to measure
obs = add(Z(i) for i in range(n_qubits))

# Initialize the QuantumModel, setting the differentiation mode
model = QuantumModel(circuit=circ, observable=obs, diff_mode=DiffMode.GPSR)

# Parameter dict with a tensor of input values for the feature parameter "phi"
# Using requires_grad=True allows differentiation w.r.t. "phi"
values={"phi": torch.rand(batch_size=1, requires_grad=True)}

# Optionally, we can also use a custom initial state, here defined as |1000>.
state = product_state("1000")

# Model execution
out_state = model.run(values=values, state=state)
samples = model.sample(values=values, state=state)
exp = model.expectation(values=values, state=state)

# Compute the gradient of the expectation value w.r.t the 
# feature parameter 'phi' using the torch.autograd API
dexp_dphi = torch.autograd.grad(exp, values["phi"], torch.ones_like(exp)) 
\end{lstlisting}
\end{shaded*}

In Code Sample \ref{lst:qm}, we showcase how to define a \lil{QuantumModel} combining a simple variational circuit and an observable to be measured. When calling model execution methods such as \lil{run}, \lil{sample}, and \lil{expectation}, a \lil{values} dictionary must be provided to map feature parameters with their concrete values. If not supplied, variational parameters values are randomly generated and passed automatically in the optimization procedure. As an example, \lil{torch.autograd} is called to compute the expectation value gradient with respect to the feature parameter \lil{phi}.

The \lil{QuantumModel} is a central object in Qadence, which inherits overridable PyTorch features for specific QML applications. For instance, the Qadence \lil{QNN} model subclasses the \lil{QuantumModel} by overriding the \lil{forward} method to calculate the observable \lil{expectation} instead of its wavefunction provided by \lil{run}. Furthermore, the \lil{QuantumModel} allows for realistic simulations by using shot-based measurement protocols or state preparation and measurement (SPAM) errors and associated error-mitigation techniques\footnote{\href{https://pasqal-io.github.io/qadence/latest/realistic_sims/}{\lil{https://pasqal-io.github.io/qadence/latest/realistic_sims/}}}.

\section{Digital-analog programming}
\label{sec:dap}

The main goal of DAQC features in Qadence is to aid users in writing digital-analog programs for a given set of interacting qubits without requiring in-depth knowledge of pulse-level specifications. For that purpose, Qadence adds background interactions to single-qubit digital gates and composes with large-scale analog operations. The following nomenclature \cite{solano2020}, referred to as \textit{strategies} later, is used throughout:
\begin{itemize}
    \item \textbf{sDAQC}: stepwise DAQC refers to programs that fully isolate the effect of single-qubit rotations, halting the effect of qubit interactions.
    \item \textbf{bDAQC}: banged DAQC refers to programs utilizing an always-on interaction, where the interaction terms still affect the system evolution during the execution of single-qubit operations.
\end{itemize}

\subsection{Translating to pre-defined Hamiltonians}

At a higher-level of abstraction and using a familiar syntax borrowed from digital quantum programming, Qadence translates digital-analog programs to a pre-defined system Hamiltonian representing Rydberg atom arrays \cite{loic2020, sebastian2023}. In these devices, atoms can be arranged in arbitrarily shaped register layouts, and computations are realized by irradiating the array with appropriately tuned laser pulses. During the computation, qubits evolve under an effective Hamiltonian,
\be{eq:ham_atoms}
    \hat{\mathcal{H}} = \sum_{i=0}^{N-1}\left(
    \frac{\Omega}{2}\left[\cos(\phi) \hat{\sigma}^x_i - \sin(\phi) \hat{\sigma}^y_i \right] 
    - \delta \hat n_i
    + \sum_{j<i}\frac{C_6}{|\mathbf r_{ij}|^6} \hat n_i \hat n_{j}\right)
\ee
where the Rabi frequency $\Omega$, detuning $\delta$ and phase $\phi$ are parameters describing global laser pulses. $\hat{n}_i = (\mathbb{1} - \hat{\sigma}^z_i)/2$ is the number operator to describe state occupancy, needed in the detuning effect and nearest neighbor interactions. The interaction strength scales with $C_6$, a coefficient dependent on the quantum number in which the atomic array has been prepared, and decays with the distance between the atoms $|r_{ij}|^6$. For more details on quantum computing with neutral atoms, we refer the reader to Ref. \cite{loic2020} as well as pulse-programming details in Ref. \cite{pulser}.

It can be seen from Eq. (\ref{eq:ham_atoms}), that finely tuning $\Omega$, $\phi$, $\delta$ and time-evolving the Hamiltonian for the appropriate duration allows for the implementation of arbitrary global $X$, $Y$ or $Z-$rotations. Turning off these parameters allows for the interaction term to freely evolve. These operations are abstracted away in dedicated Qadence blocks shown in Code Sample \ref{lst:da-prog}.

\begin{shaded*}
\begin{lstlisting}[caption=Digital-analog convenience operators, label={lst:da-prog}, language=python]
# Global rotations automatically translated to the Hamiltonian parameters
rx, ry, rz = AnalogRX(angle="th1"), AnalogRY(angle="th2"), AnalogRZ(angle="th3")

# Evolve the interaction term
analog_int = AnalogInteraction(duration = "t")

# Fully control all the parameters
da_rot = AnalogRot(omega="om", phase="ph", delta="d", duration="t")
\end{lstlisting} 
\end{shaded*}

\noindent Note that parameters passed as strings are converted into instances of \lil{VariationalParameter}. By definition, rotations in $\hat{\mathcal{H}}$ are globally applied to the full qubit support, in accordance with the analog mode in Rydberg
atom arrays. Currently, this is done within the bDAQC strategy by implicitly adding background interaction to digital gates, but will soon be customizable to the sDAQC strategy. This is supported in two backends:
\begin{itemize}
    \item Pulser, where respective operations are approximated by square pulses. This creates a direct connection between high-level digital-analog programs and pulse sequences for execution on real devices.
    \item PyQTorch, where respective operations are converted to Hamiltonian matrices and exponentiated in the statevector simulator, allowing for more efficient prototyping of simple programs.
\end{itemize}

\subsection{Flexible Hamiltonian construction}

For users familiar with programming Hamiltonian operations, Qadence provides a convenient \lil{hamiltonian_factory} to quickly build Hamiltonians composed from Pauli operators. For a given register with a topology graph $G(V,E)$, it produces a Hamiltonian of the type:
\be{eq:ham_factory}
    \hat{\mathcal{H}}=\sum_{i\in V}\alpha_i\hat O_i+\sum_{(i,j)\in E}\beta_{ij}\hat{\mathcal{H}}^\text{int}_{ij}
\ee
where the single-qubit detuning operator $\hat{O}_i\in\{\hat{\sigma}^x,\hat{\sigma}^y,\hat{\sigma}^z,\hat{n}\}$ can be chosen with the \lil{detuning} argument, and the interaction $\hat{\mathcal{H}}^\text{int}$ can be chosen from the \lil{Interaction} enumeration as \lil{NN}, \lil{ZZ}, \lil{XY} or \lil{XYZ}, or provided as a custom function. The detuning and interaction strengths $\alpha_i$ and $\beta_{ij}$ can be set to any supported parameter type. Interactions can also be customized for the complete set of node pairs instead of the edges in the topology graph, as exemplified below.

\begin{shaded*}
\begin{lstlisting}[caption=Arbitrary Hamiltonians, label={lst:arb-ham}, language=python]
reg = Register.triangular_lattice(n_cells_row=2, n_cells_col=2, spacing=2.0)

# Create the interaction strength term with 1/r decay
strength_list = [1.0 / reg.distances[p] for p in reg.all_node_pairs]

# Initialize NN Hamiltonian
nn_ham = hamiltonian_factory(
    reg,                                # Register with the Hamiltonian topology
    interaction=Interaction.NN,         # Type of interaction to use
    interaction_strength=strength_list, # List of all interaction strengths
    detuning=X,                         # Pauli operator for the detuning
    detuning_strength="d",              # Parameterize the detuning strength
    use_all_node_pairs=True,            # Use all pairs instead of graph edges
)
\end{lstlisting} 
\end{shaded*}

In order to provide a more generic Hamiltonian oriented programming experience the interaction can also be a user-defined function signed with two integers.

\begin{shaded*}
\begin{lstlisting}[caption=User-defined Hamiltonians, label={lst:user-ham}, language=python]
def custom_interaction(i: int, j: int):
    return X(i) @ X(j) + Y(i) @ Y(j)
    
custom_ham = hamiltonian_factory(reg, interaction = custom_interaction)
\end{lstlisting}
\end{shaded*}

\subsection{Hamiltonian transformation and other digital-analog constructors}

Another important tool in DAQC programs is the ability to map, or transform, the evolution of Ising Hamiltonians\footnote{In this case, we are referring specifically to Hamiltonians composed of $ZZ$- or $NN$-interaction types, which so far we have written with the notation $\hat{\sigma}^z_i\hat{\sigma}^z_j$ and $\hat{n}_i\hat{n}_j$.} one into another, through the universality property of more general entangling Hamiltonians \cite{dodd2002, solano2020}. Therefore, for a given sequence of unitary operations of interest, described as the evolution of a some \textit{target} Hamiltonian, the DAQC transform provides a unique mapping to the evolution of a fixed system Hamiltonian (the \textit{build} Hamiltonian) with extra single-qubit rotations. This transform increases the flexibility and usability of interacting qubit systems as in Eq. (\ref{eq:ham_atoms}) for instance, by allowing the implementation of algorithms that would otherwise require considerable physical rearrangements. In Qadence, this transformation is readily available with the \lil{daqc_transform} function for entangling Hamiltonians with $ZZ$ and $NN$-interaction types, the latter requiring extra local detunings (single qubit rotations) and an optional global phase. As an example below, we show how we can obtain the circuit that evolves a given local target Hamiltonian for some fixed time using only the evolution of the triangular lattice Hamiltonian from Code Sample \ref{lst:arb-ham} and local single-qubit rotations.


\begin{shaded*}
\begin{lstlisting}[caption=Hamiltonian transformation, label={lst:ham-map}, language=python]
# Target Hamiltonian to evolve for a specific time t_f
h_target = N(0)@N(1) + N(1)@N(2) + N(2)@N(0)
t_f = 5.0

transformed_ising = daqc_transform(
    n_qubits = reg.n_qubits,
    gen_target = h_target,
    t_f = t_f,
    gen_build = nn_ham,
    strategy=Strategy.SDAQC,
)
\end{lstlisting} 
\end{shaded*}

\noindent Currently, only the sDAQC strategy is supported in \lil{daqc_transform}, with a planned extension to bDAQC soon. As an example of a DAQC-transformed algorithm, the digital-analog QFT \cite{daqft2020} uses the intrinsic Ising structure of each layer of CPHASE gates in the digital QFT to apply the previously described transformation. The DA-QFT is available in Qadence by calling the \lil{qft} constructor and passing the respective strategy and desired build Hamiltonian.
\begin{shaded*}
\begin{lstlisting}[caption=Digital-Analog QFT, label={lst:da-qft}, language=python]
# Changing the strategy overrides the default Strategy.DIGITAL
qft_daqc = qft(n_qubits=3, strategy=Strategy.SDAQC, gen_build=h_build)
\end{lstlisting} 
\end{shaded*}

\section{Applications}\label{sec:apps}

In this Section, we present two paradigmatic applications that leverage the full flexibility of the Qadence interface for building digital-analog variational quantum algorithms. We showcase two types of usecases:

\begin{itemize}
    \item \textit{Quantum Machine Learning} (QML) - Solving differential equations using quantum neural networks. We show an implementation of the Differentiable Quantum Circuit (\textbf{DQC}) algorithm \cite{dqc1}.
    \item \textit{Combinatorial Optimization} - Solving a quadratic unconstrained optimization (QUBO) problem. We show an implementation of the variational \textbf{QAOA} algorithm \cite{qaoa2014} using a fully analog circuit.
\end{itemize}

\subsection{QML: Solving differential equations using DQC}\label{sec:apps-dqc}

To illustrate Qadence's ability to define and train QML models, we implement a DQC model that will then be used to solve the following non-linear ordinary differential equation (ODE) 
\be{label:diff_eq}
\frac{df}{dx}=4x^3+x^2-2x-\frac{1}{2}
\ee
with a Dirichlet boundary condition $f(0) = 1$, and a known closed-form solution: $f(x) = x^4+\frac{1}{3}x^3 - x^2 - \frac{1}{2}x+1$. 

\begin{figure}[t]
    \centering
    \includegraphics[width=0.55\linewidth]{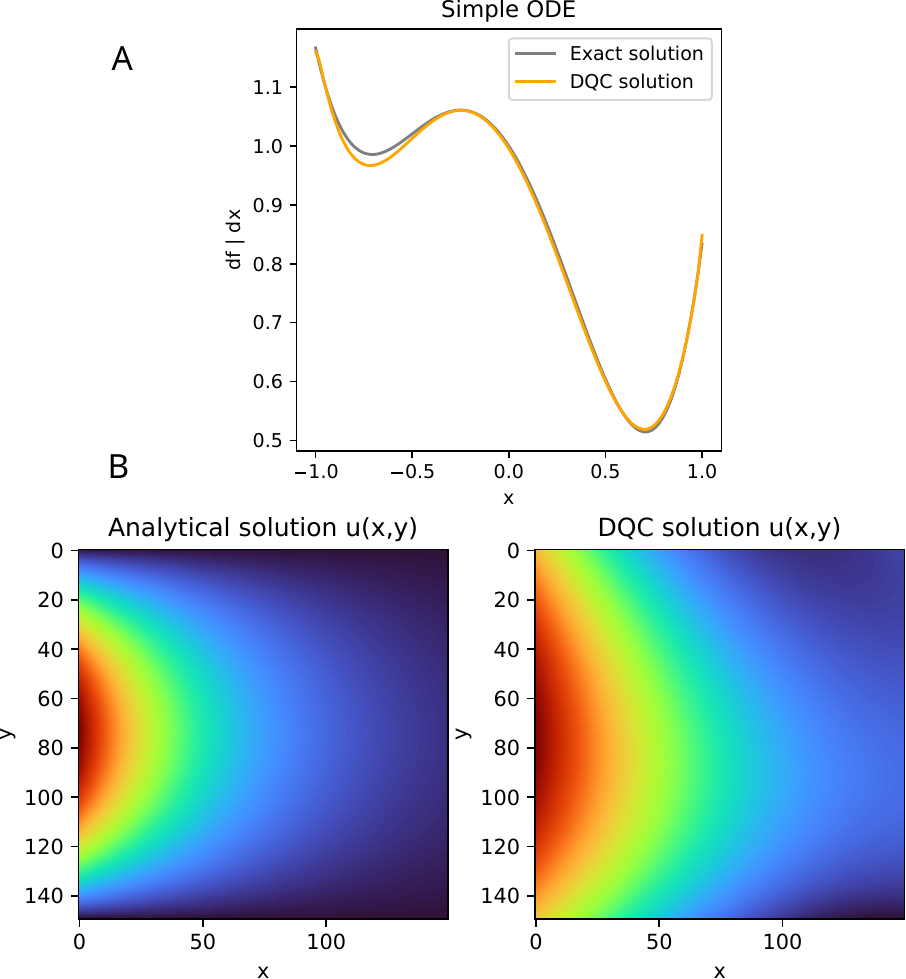}
    \caption{A) Solution of the simple non-linear ODE presented
    in the text using DQC. The solution is computed within the 
    $(-1,1)$ domain for $x$ and the results are compared to the 
    known closed-form solution after gradient-based training with a random uniform sample of 20 collocation points. B) Solution of 2D Laplace equation presented
    in the text using DQC. The solution is computed within the 
    $[0,1]$ domains for both $x$ and $y$ and the results are compared to the 
    known closed-form solution after gradient-based training with a random uniform sample of 100 collocation points. For both examples, 1000 epochs of training with the Adam optimizer and a learning rate of 0.01 was used. The DQC comprised a hardware-efficient ansatz with 4 qubits and depth = 3, a Chebyshev feature map for data encoding, and a transverse-field Ising Hamiltonian as a cost function.
    }
    \label{fig:dqc_fig}
\end{figure}

Qadence simplifies the QML model building by providing components for classical data embedding through pre-defined or custom-made feature maps, ans{\"a}tze with flexible architectures, and model training tools. To solve this simple ODE we make use of the following ingredients:
\begin{itemize}
    \item a digital hardware-efficient ansatz (HEA) \cite{hea}.
    \item a Chebyshev feature map (see Ref. \cite{dqc1} for more information).
    \item a transverse-field Ising Hamiltonian as a measured observable.
    \item a standard PyTorch Adam optimizer \cite{Kingma2014}.
\end{itemize}
The essential part of solving this problem with Qadence is to define a problem-specific loss function that incorporates the physical constraints as regularisation terms of the loss during training. In the code sample below we exemplify all of these steps, while the complete implementation is provided in Appendix \ref{app:dqc_ode}. The solution is shown in Fig.~\ref{fig:dqc_fig}.

Qadence also provides the user with the flexibility to easily define and train more complex QML models, and solve realistic partial differential equations (PDEs). We show this by solving the 2D Laplace equation with a DQC model. The Laplace equation is a PDE with the following form:
\be{label:laplace_eq}
\frac{\partial^2u}{\partial x^2} + \frac{\partial^2u}{\partial y^2}=0
\ee
and it has a known exact solution $u(x,y)=e^{-\pi x}sin(\pi y)$ when solved with the Dirichlet boundary conditions $u(0,y)=sin(\pi y)$, $u(x,0)=0$, $u(X,y)=0$, $u(x,Y)=0$ solved for $x,y \in [0, 1]$. 
The DQC solution is shown in Fig.~\ref{fig:dqc_fig} and a fully functional implementation is available in Appendix \ref{appendix:pde}.

\subsection{Combinatorial Optimization: Solving a QUBO problem using QAOA}\label{sec:apps-qubo}

Qadence can not only be used in QML contexts but is suitable for any kind of variational algorithm, especially if digital-analog or purely analog computations are required. We show next how to solve a QUBO problem using the QAOA variational algorithm. QUBOs are combinatorial optimization problems with a wide range of applications \cite{qubo}. A QUBO function can be represented in the following way:
\begin{equation} \label{eq:qubo}
    f_Q(z) = \sum_{i,j}^N Q_{ij} z_i z_j = Z^{\textrm{T}} Q Z,
\end{equation}
where $z_i$ are binary variables arranged in a vector $Z = (z_1, \dots, z_N)$ and $Q$ is a symmetric matrix defining the coefficients of the binary vector. The optimization procedure aims at finding the set of binary variables that yields the lowest value of the function $f_Q(z)$. QUBO problems are NP-hard and the solution might not be unique. It has been shown \cite{nguyen2023} that QUBO problems can be encoded onto Rydberg atom arrays and solved either adiabatically \cite{farhi2000quantum} or via a variational approach based on the QAOA algorithm \cite{qaoa2014}. Here we use the latter. For solving a QUBO, we require the following components:
\begin{itemize}
    \item A symmetric, real-valued matrix $Q$ representing the QUBO weight coefficients.
    \item A suitable loss function that computes Eq. (\ref{eq:qubo}) from a given measurement.
    \item A register with a specific spatial arrangement of the Rydberg atoms that embeds the QUBO problem in the interaction Hamiltonian. Finding this arrangement is problem- and hardware-dependent, and discussing it is beyond the scope of this manuscript (see Ref. \cite{nguyen2023} for details).
    \item An ansatz, chosen to be a series of analog quantum operations with parametrized angles.
    \item A gradient-free optimizer for sample-based optimization of the variational circuit.
\end{itemize}
Once the QUBO problem is embedded in the atomic register, the variational angles are optimized during the QAOA procedure and we check if the system converges to the optimal solution. All these components are shown in Code Sample \ref{lst:qubo-loss}. 

\begin{shaded*}
\begin{lstlisting}[caption=Solve a QUBO problem with an analog program., label={lst:qubo-loss}, language=python]
# Below we consider a symmetric matrix Q encoding the QUBO problem
# And a function that translates Q into the necessary qubit coordinates
reg = Register.from_coordinates(qubo_register_coords(Q))

# Fully anaog ansatz composed of global X and Z rotations
ansatz = chain(*[AnalogRX(f"t{i}") * AnalogRZ(f"s{i}") for i in range(N_LAYERS)])
model = QuantumModel(QuantumCircuit(reg, ansatz))

optimizer = ... #  A gradient-free optimizer using Nevergrad

def loss_fn(model: QuantumModel, *args) -> tuple[float, dict]:
    to_array = lambda bitstring: np.array(list(bitstring), dtype=int) 
    cost_fn = lambda Z: Z.T @ Q @ Z
    samples = model.sample({}, n_shots=N_SHOTS)[0] 
    cost = sum(samples[key] * cost_fn(to_array(key)) for key in samples)
    return cost / N_SHOTS, {}

train_no_grad(model, None, optimizer, config, loss_fn)
\end{lstlisting} 
\end{shaded*}

\begin{figure}[t]
    \centering
    \includegraphics[width=0.8\linewidth]{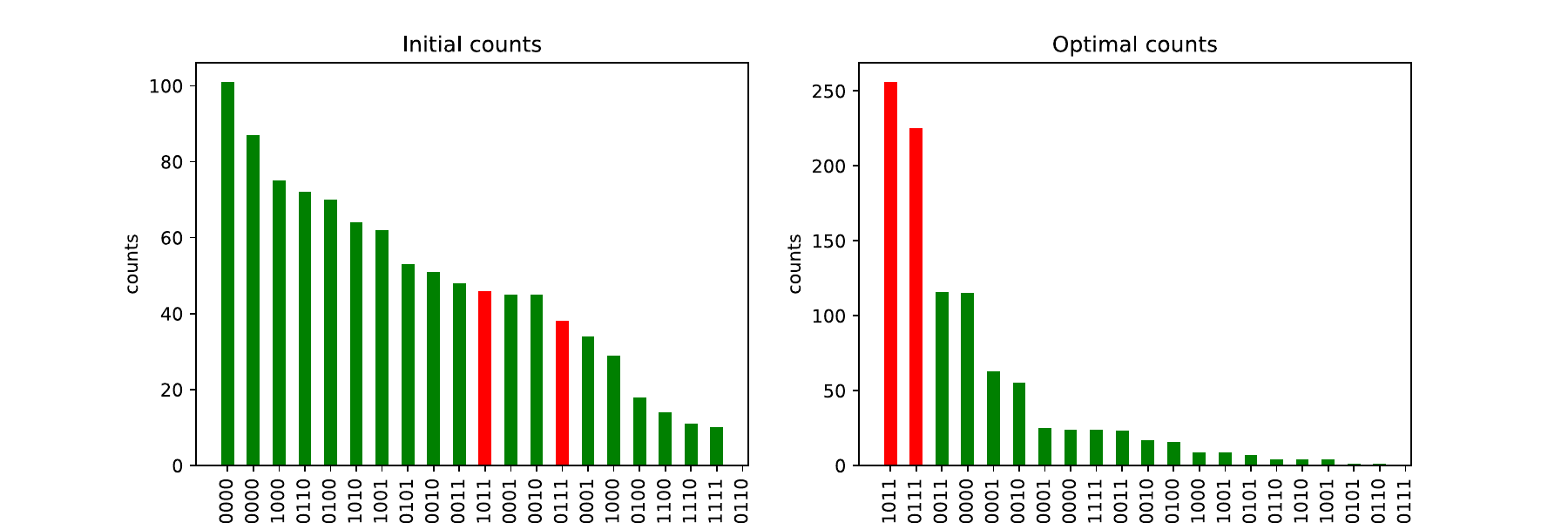}
    \caption{QUBO solutions before optimization (left panel) and
    after optimization (right panel). The red bars are the expected
    solutions of the problem.}
    \label{fig:qubo-solution}
\end{figure}

\noindent Plotting the initial and optimal counts results in Fig.~\ref{fig:qubo-solution} where the red bars correspond to the expected optimal (equivalent) solutions of the QUBO problem. For the complete working code, the reader can refer to Appendix \ref{appendix:qubo}.

\section{Conclusions}\label{sec:concl}

In this technical report, we presented Qadence, a high-level 
programming interface for creating and running digital-analog 
quantum computing programs on various numerical and physical backends. Qadence programs are executable on emulators and platforms compatible with
the DAQC paradigm, such as Rydberg neutral atom arrays \cite{loic2020}.

Qadence offers a flexible block-based interface that enables users
to write abstract DAQC programs with a language very close to the
mathematical representation of the circuit operations, including the
complex Hamiltonian evolution unitaries needed by the DAQC
paradigm, and qubit interaction
represented on a coordinate-based register with configurable
connectivity. Variational parameters with arbitrary expressions 
can also be assigned in a straightforward manner using 
symbolic algebraic expressions based on the popular \lil{Sympy}
library \cite{sympy}.

Qadence backends include
a fast and efficient PyTorch-based quantum statevector simulator \cite{pyqtorch},
a numerical simulator based on tensor networks \cite{anton_2024_emuc}
(to be open-sourced soon) and real devices such as Pasqal's Rydberg atom arrays accessible via the 
Pulser \cite{pulser} backend.

Since Qadence is tailored for variational algorithm research 
and, particularly, quantum machine learning applications, significant
attention has been devoted to ensuring efficient quantum circuit
differentiability on all these backends.
This is achieved numerically by integrating with popular deep-learning
frameworks whose 
AD engine can be fully leveraged in
Qadence programs and experimentally with device amenable GPSR \cite{gpsr2021}. 


As shown in the previous section, Qadence enables users 
to write variational applications with
few lines of code whilst keeping great readability and clarity. We believe that this library fills an important
gap in the open-source quantum computing software space and hope it will become the \emph{de facto}
standard for executing DAQC quantum programs in the future.

\section{Future Work}
\label{sec:future}

Qadence is currently under active development, and future efforts aim at modularizing core language features from an ecosystem of additional domain-specific libraries for user-friendliness and rapid uptake. The objective is for Qadence to become a versatile middleware cornerstone in a software execution stack for DAQC hardware providers as well as an efficient tool for DAQC algorithmic development and applications.
Fundamental developments will concern robustness and expressivity of the block system  and the ability to run noisy simulations at scale. A leaner and more intuitive high-level digital-analog interface is under development with a planned possibility to customize the interacting Hamiltonian beyond the currently supported one for Rydberg atom arrays. Performing digital-analog transformations with using the bDAQC strategy is also currently under development \cite{solano2020}, and also the possibility of using fully arbitrary resource and target Hamiltonians \cite{daqc-arbitrary}. Dependencies externalisation will enable realistic simulations with so-called \textit{protocols} for efficient shot-based measurements and hardware amenable error mitigation techniques. Resource efficiency for hardware execution will also be prioritized by enabling a compilation step together with circuit and pulse optimization passes. Performance and higher-order differentiability will be tackled in forthcoming releases of numerical simulators and emulators integrated with Qadence.

\bibliographystyle{unsrt}
\bibliography{references}

\appendix

\section{Differentiation using the low-level API}
\label{appendix:diff}

In this Appendix, we show a working example of how to
use the low-level backend API of Qadence and select the
desired differentiation mode. Note that this low-level interface is not intended to be exposed to the user but shown here for the sake of example.

\begin{shaded*}
\begin{lstlisting}[caption=Differentiation modes for quantum circuits, label={lst:diff_mode}, language=python]
from qadence import kron, RX, FeatureParameter, total_magnetization, QuantumCircuit
from qadence.backends import backend_factory
import torch
import sympy

def differentiate(diff_mode, circuit, observable, values):

    # instantiate a differentiable backend with the given differentiation
    # mode using the PyQTorch statevector simulator backend
    backend = backend_factory(backend="PYQTORCH", diff_mode=diff_mode)
    
    # convert circuit, observable and circuit parameters
    # to a representation suitable for the chosen DifferentiableBackend object
    converted = backend.convert(circuit, observable)
    embedded_params = converted.embedding_fn(converted.params, values)
    
    # compute the expectation value and differentiate it with respect
    # to the "x" parameter using the standard torch.autograd engine
    expval = backend.expectation(
        converted.circuit, converted.observable, param_values=embedded_params
    )
    return torch.autograd.grad(expval, values["x"], torch.ones_like(expval))[0] 

x = FeatureParameter("x")
circuit = QuantumCircuit(n_qubits, kron(RX(i, (i+1) * sympy.acos(x)) for i in range(4)))
observable = total_magnetization(4)

values = {"x": torch.rand(10, requires_grad=True)}
diff_ad = differentiate("AD", circuit, observable, values)
diff_gpsr = differentiate("GPSR", circuit, observable, values)
diff_adjoint = differentiate("ADJOINT", circuit, observable, values)

# check that derivatives are matching
check_eq = lambda x, y : torch.all(torch.isclose(x, y)).item()
assert check_eq(diff_ad, diff_gpsr) and check_eq(diff_gpsr, diff_adjoint)
\end{lstlisting}
\end{shaded*}

\section{Complete DQC example solving an ODE}\label{app:dqc_ode}

Here, we show the full implementation of the \textbf{QML} example of Sec. \ref{sec:apps-dqc}.

\begin{shaded*}
\begin{lstlisting}[caption=Solving a simple non-linear ODE with DQC., label={lst:dqc_simple_ode}, language=python]
# library imports
import matplotlib.pyplot as plt
import numpy as np
from numpy.random import uniform
from qadence import *
from torch import nn, optim, tensor, zeros_like, ones_like, linspace, manual_seed
from torch.autograd import grad

#  random seed
manual_seed(404)

N_QUBITS, DEPTH, LEARNING_RATE, N_POINTS = 4, 3, 0.01, 20

# building the DQC model
ansatz = hea(n_qubits=N_QUBITS, depth=DEPTH)
# the input data is encoded via a feature map
fm = feature_map(n_qubits=N_QUBITS, param="x", fm_type="chebyshev")
# choosing a cost function
obs = ising_hamiltonian(n_qubits=N_QUBITS)
# building the circuit and the quantum model
circuit = QuantumCircuit(N_QUBITS, chain(fm, ansatz))
model = QNN(circuit=circuit, observable=obs, inputs=["x"])

# using Adam as an optimizer of choice
opt = optim.Adam(model.parameters(), lr=LEARNING_RATE)

# define a problem-specific MSE loss function
# for the ODE df/dx=4x^3+x^2-2x-1/2
def loss_fn(inputs: Tensor, outputs: Tensor) -> Tensor:
    dfdx = grad(inputs=inputs, outputs=outputs.sum(), create_graph=True)[0]
    ode_loss = dfdx - (4 * inputs**3 + inputs**2 - 2 * inputs - 0.5)
    boundary_loss = model(zeros_like(inputs)) - ones_like(inputs)
    return ode_loss.pow(2).mean() + boundary_loss.pow(2).mean()

# collocation points sampling and training
for epoch in range(1000):
    opt.zero_grad()
    # the collocation points are sampled randomly
    cp = tensor(
        uniform(low=-0.99, high=0.99, size=(N_POINTS, 1)), requires_grad=True
    ).float()
    loss = loss_fn(inputs=cp, outputs=model(cp))
    loss.backward()
    opt.step()

# compare the solution to known ground truth

sample_points = linspace(-1.0, 1.0, steps=100).reshape(-1, 1)
# analytical solution
analytic_sol = (
    sample_points**4
    + (1 / 3) * sample_points**3
    - sample_points**2
    - (1 / 2) * sample_points
    + 1
)

# DQC solution
dqc_sol = model(sample_points).detach().numpy()

x_data = sample_points.detach().numpy()

# plot

plt.figure(figsize=(4, 4))
plt.plot(x_data, analytic_sol.flatten(), color="gray", label="Exact solution")
plt.plot(x_data, dqc_sol.flatten(), color="orange", label="DQC solution")
plt.xlabel("x")
plt.ylabel("df | dx")
plt.title("Simple ODE")
plt.legend()
plt.show()
\end{lstlisting}
\end{shaded*}

\section{Complete DQC example solving a PDE}
\label{appendix:pde}

Here we provide the DQC code used to solve the 2D Laplace example showed in Sec~\ref{sec:apps-dqc} of the main text.

\begin{shaded*}
\begin{lstlisting}[caption=Complete DQC example solving a PDE, label={lst:pde-complete}, language=python]
# library imports
from itertools import product
import matplotlib.pyplot as plt
import numpy as np
from qadence import *
from torch import (
    nn,
    optim,
    tensor,
    ones,
    zeros,
    zeros_like,
    ones_like,
    sin,
    exp,
    rand,
    linspace,
    manual_seed,
)
from torch.autograd import grad

#  random seed
manual_seed(42)

# helper function to calculate derivatives
def calc_derivative(outputs, inputs) -> tensor:
    """
    Returns the derivative of a function output
    with respect to its inputs
    """
    if not inputs.requires_grad:
        inputs.requires_grad = True
    return grad(
        inputs=inputs,
        outputs=outputs,
        grad_outputs=ones_like(outputs),
        create_graph=True,
        retain_graph=True,
    )[0]


class DomainSampling(nn.Module):
    """
    Collocation points sampling from domains uses uniform random sampling.
    Problem-specific MSE loss function for solving the 2D Laplace equation.
    """

    def __init__(self, net: nn.Module | QNN, n_inputs: int = 2, n_colpoints: int = 20):
        super().__init__()
        self.net = net
        self.n_colpoints = n_colpoints
        self.n_inputs = n_inputs

    def left_boundary(self) -> tensor:  # u(0,y)=0
        sample = rand(size=(self.n_colpoints, self.n_inputs))
        sample[:, 0] = 0.0
        return self.net(sample).pow(2).mean()

    def right_boundary(self) -> tensor:  # u(L,y)=0
        sample = rand(size=(self.n_colpoints, self.n_inputs))
        sample[:, 0] = 1.0
        return self.net(sample).pow(2).mean()

    def top_boundary(self) -> tensor:  # u(x,H)=0
        sample = rand(size=(self.n_colpoints, self.n_inputs))
        sample[:, 1] = 1.0
        return self.net(sample).pow(2).mean()

    def bottom_boundary(self) -> tensor:  # u(x,0)=f(x)
        sample = rand(size=(self.n_colpoints, self.n_inputs))
        sample[:, 1] = 0.0
        return (self.net(sample) - sin(np.pi * sample[:, 0])).pow(2).mean()

    def interior(self) -> tensor:  # uxx+uyy=0
        sample = rand(size=(self.n_colpoints, self.n_inputs), requires_grad=True)
        first_both = calc_derivative(self.net(sample), sample)
        second_both = calc_derivative(first_both, sample)
        return (second_both[:, 0] + second_both[:, 1]).pow(2).mean()


LEARNING_RATE = 0.01 
N_QUBITS = 4
DEPTH = 3
VARIABLES = ("x", "y")
N_POINTS = 150

# define a simple DQC model
ansatz = hea(n_qubits=N_QUBITS, depth=DEPTH)
# parallel Fourier feature map
split = N_QUBITS // len(VARIABLES)
fm = kron(
    *[
        feature_map(n_qubits=split, support=support, param=param)
        for param, support in zip(
            VARIABLES,
            [
                list(list(range(N_QUBITS))[i : i + split])
                for i in range(N_QUBITS)
                if i % split == 0
            ],
        )
    ]
)
# choosing a cost function
obs = ising_hamiltonian(n_qubits=N_QUBITS)

# building the circuit and the quantum model
circuit = QuantumCircuit(N_QUBITS, chain(fm, ansatz))
model = QNN(circuit=circuit, observable=obs, inputs=VARIABLES)

# using Adam as an optimiser of choice
opt = optim.Adam(model.parameters(), lr=LEARNING_RATE)

# get the collocation sampling for loss calculation
sol = DomainSampling(net=model, n_inputs=2, n_colpoints=100)

# training
for epoch in range(1000):
    opt.zero_grad()
    loss = (
        sol.left_boundary()
        + sol.right_boundary()
        + sol.top_boundary()
        + sol.bottom_boundary()
        + sol.interior()
    )
    loss.backward()
    opt.step()

# visualisation and comparison of results

single_domain = linspace(0, 1, steps=N_POINTS)
domain = tensor(list(product(single_domain, single_domain)))
# analytical solution
analytic_sol = (
    (exp(-np.pi * domain[:, 0]) * sin(np.pi * domain[:, 1]))
    .reshape(N_POINTS, N_POINTS)
    .T
)

# DQC solution
dqc_sol = model(domain).reshape(N_POINTS, N_POINTS).detach().numpy()

# plot results

fig, ax = plt.subplots(1, 2, figsize=(7, 7)

ax[0].imshow(analytic_sol, cmap="turbo")
ax[0].set_xlabel("x")
ax[0].set_ylabel("y")
ax[0].set_title("Analytical solution u(x,y)")

ax[1].imshow(dqc_sol, cmap="turbo")
ax[1].set_xlabel("x")
ax[1].set_ylabel("y")
ax[1].set_title("DQC solution u(x,y)")

plt.show()
\end{lstlisting}
\end{shaded*}

\section{Complete QUBO example}
\label{appendix:qubo}

In this Appendix, we show a complete working example of the QUBO solver
presented in Sec~\ref{sec:apps} of the main text.

\begin{shaded*}
\begin{lstlisting}[caption=Complete QUBO example, label={lst:qubo-complete}, language=python]
from __future__ import annotations

from typing import Any

import matplotlib.pyplot as plt
import nevergrad as ng
import numpy as np
import torch
from scipy.optimize import minimize
from scipy.spatial.distance import pdist, squareform

from qadence import AnalogRX, AnalogRZ, QuantumCircuit, QuantumModel, Register, RydbergDevice, chain
from qadence.analog.constants import C6_DICT
from qadence.ml_tools import TrainConfig, num_parameters, train_gradient_free

# Setting seeds for reproducibility
seed = 0
np.random.seed(seed)
torch.manual_seed(seed)


def qubo_register_coords(Q: np.ndarray) -> list[tuple[int, int]]:
    """Compute coordinates for register."""

    def evaluate_mapping(new_coords: np.ndarray, *args: Any) -> Any:
        """Cost function to minimize. Ideally, the pairwise.

        distances are conserved
        """
        Q, shape = args
        new_coords = np.reshape(new_coords, shape)
        rydberg_level = 70
        interaction_coeff = C6_DICT[rydberg_level]
        new_Q = squareform(interaction_coeff / pdist(new_coords) ** 6)
        return np.linalg.norm(new_Q - Q)

    shape = (len(Q), 2)
    np.random.seed(0)
    x0 = np.random.random(shape).flatten()
    res = minimize(
        evaluate_mapping,
        x0,
        args=(Q, shape),
        method="Nelder-Mead",
        tol=1e-6,
        options={"maxiter": 200000, "maxfev": None},
    )
    return [(x, y) for (x, y) in np.reshape(res.x, (len(Q), 2))]


# The number of times we want to sample from our candidate model
N_SHOTS = 1000

# QUBO problem weights, a real-valued, symmetric matrix.
Q = np.array(
    [
        [-10.0, 19.7365809, 19.7365809, 5.42015853, 5.42015853],
        [19.7365809, -10.0, 20.67626392, 0.17675796, 0.85604541],
        [19.7365809, 20.67626392, -10.0, 0.85604541, 0.17675796],
        [5.42015853, 0.17675796, 0.85604541, -10.0, 0.32306662],
        [5.42015853, 0.85604541, 0.17675796, 0.32306662, -10.0],
    ]
)


# QUBO loss function
def loss_fn(model: QuantumModel, *args: Any) -> tuple[float, dict[str, float]]:
    to_array = lambda bitstring: np.array(
        list(bitstring), dtype=int
    )  # Convert a bitstring to a np.array
    cost_fn = lambda Z: Z.T @ Q @ Z  # Compute its cost given a fixed matrix Q
    samples = model.sample({}, n_shots=N_SHOTS)[
        0
    ]  # Sample from the model given current parameter values
    cost = sum(
        samples[key] * cost_fn(to_array(key)) for key in samples
    )  # Compute the cost for each sampled bitstring and weight it by its sample frequency
    return cost / N_SHOTS, {}  # Normalize the cost by the number of samples
    # We return an optional metrics dict


# Initialize the register with Rydberg device specifications
device = RydbergDevice(rydberg_level=70)

reg = Register.from_coordinates(qubo_register_coords(Q), device_specs=device)

# Analog circuit using Rydberg atom interaction for solving the QUBO.
# Each layer is parametrized with the evolution time of the two analog rotations
n_layers = 2
ansatz = chain(*[AnalogRX(f"t{i}") * AnalogRZ(f"s{i}") for i in range(n_layers)])
model = QuantumModel(QuantumCircuit(reg, ansatz), backend="pyqtorch")

initial_counts = model.sample({}, n_shots=1000)[0]
# Define a config which tells qadence how many epochs we want to train our model
config = TrainConfig(max_iter=100)
# Our loss is computed using samples which is a non-differentiable operation
# Hence we opt for a gradient-free optimizer
optimizer = ng.optimizers.NGOpt(budget=config.max_iter, parametrization=num_parameters(model))

# qadence.ml_tools offers a training routine for gradient-free optimization tasks
data_loader = None
# Since we do not require external data, we pass a 'None' as data_loader
train_gradient_free(model, data_loader, optimizer, config, loss_fn)

# Sample from the model using the optimal values for our variational parameters
optimal_counts = model.sample({}, n_shots=1000)[0]

def plot_distribution(C: dict[str, int], ax: Any, title: str) -> None:
    C = dict(sorted(C.items(), key=lambda item: item[1], reverse=True))
    indexes = ["01011", "00111"]  # QUBO solutions
    color_dict = {key: "r" if key in indexes else "g" for key in C}
    ax.set_xlabel("bitstrings")
    ax.set_ylabel("counts")
    ax.set_xticks([i for i in range(len(C.keys()))], C.keys(), rotation=90)
    ax.bar(C.keys(), C.values(), width=0.5, color=color_dict.values())
    ax.set_title(title)


fig, axs = plt.subplots(1, 2, figsize=(12, 4))
plot_distribution(initial_counts, axs[0], "Initial counts")
plot_distribution(optimal_counts, axs[1], "Optimal counts")
plt.show()
\end{lstlisting}
\end{shaded*}

\end{document}